\documentclass[aps,pra,10pt,twocolumn,superscriptaddress,floatfix,nofootinbib,showpacs,longbibliography]{revtex4-2}

\usepackage{graphicx}
\usepackage{amsmath,amssymb,amsthm}
\usepackage{xcolor}
\usepackage{braket}
\usepackage{natbib}
\usepackage[normalem]{ulem}
\usepackage{tikz}
\usepackage{quantikz}
\usepackage{comment}
\usepackage{hyperref}
\usepackage{array}
\usepackage{lmodern}
\usepackage[T1]{fontenc}
\usepackage{booktabs}
\usepackage{floatrow}

\usepackage[ruled,vlined,noline]{algorithm2e}
\SetAlgoSkip{smallskip}

\makeatletter
\renewcommand{\algocf@caption@ruled}{%
  \box\algocf@capbox\vskip.6\baselineskip}
\makeatother
\begin{document}

\title{Deterministic Quantum Phase Estimation with Linear \\ Circuit Complexity in a Photonic System}

\author{M. Midhuna}
\thanks{These authors contributed equally to this work.}
\affiliation{%
Quantum Optics \& Quantum Information, Department of Electronic Systems Engineering, Indian Institute of Science, Bengaluru 560012, India
}
\author{Ajay Jayachandran}
\thanks{These authors contributed equally to this work.}
\affiliation{%
Quantum Optics \& Quantum Information, Department of Electronic Systems Engineering, Indian Institute of Science, Bengaluru 560012, India
}
\author{Kanad Sengupta}
\affiliation{%
Quantum Optics \& Quantum Information, Department of Electronic Systems Engineering, Indian Institute of Science, Bengaluru 560012, India
}
\affiliation{%
 Department of Instrumentation and Applied Physics, Indian Institute of Science, Bengaluru 560012, India}

\author{Akshai T. Krishnan}
\affiliation{%
Quantum Optics \& Quantum Information, Department of Electronic Systems Engineering, Indian Institute of Science, Bengaluru 560012, India
}
\affiliation{%
 Department of Instrumentation and Applied Physics, Indian Institute of Science, Bengaluru 560012, India}

\author{C. M. Chandrashekar}

\affiliation{%
Quantum Optics \& Quantum Information, Department of Electronic Systems Engineering, Indian Institute of Science, Bengaluru 560012, India
}


\begin{abstract}

Quantum algorithms solve certain computational problems faster than the best known classical algorithms. Many algorithms, including Shor’s for factoring, Grover’s for unstructured search, and the HHL for solving linear systems, rely on quantum phase estimation (QPE) as a fundamental subroutine. The QPE protocol proceeds through the initialization of a control register in a uniform superposition, controlled unitary evolution encoding the eigenphase, and a final inverse quantum Fourier transform followed by measurement to extract the phase. Here, we address a special class of unitary operators that frequently appear in quantum Fourier transform-based protocols, cyclic group representations, and periodically evolving quantum systems. We introduce a QPE algorithm that successfully reduces the circuit complexity from $\mathcal{O}(n^2)$ to $\mathcal{O}(n)$ for special class of unitary operators and implement it on a four-qubit photonic system.  The four-qubit system is realised using a photon pair, with two qubits encoded in its polarization degree of freedom and the remaining two in its path modes. In contrast to previous photonic implementations of QPE based on dual-rail encoding and KLM protocol, where controlled operations are inherently probabilistic and thus reduce the overall success probability of phase estimation, our scheme is fully deterministic. Moreover, it is scalable to higher-dimensional unitaries, provided the underlying structure of the unitaries is preserved.
\end{abstract}

  \maketitle


\section{Introduction}

Quantum phase estimation (QPE) is one of the most fundamental algorithms in quantum computing and forms the basis of numerous quantum algorithms that offer exponential or polynomial speedups over their classical counterparts \cite{Shor1994,Harrow2009,Brassard2002quantum,Nielsen2000,abrams1999quantum,brassard1998quantum}. Since its introduction, QPE has become an indispensable subroutine in a wide range of applications, including Shor's factoring algorithm, the Harrow--Hassidim--Lloyd (HHL) algorithm for solving systems of linear equations, quantum simulation, quantum amplitude estimation, Hamiltonian eigenvalue estimation, quantum principal component analysis, and quantum algorithms for data fitting and machine learning \cite{Shor1994,Harrow2009,Brassard2002quantum,Nielsen2000,lloyd2014quantum,wiebe2012quantum,lloyd2013quantum}. At its core, the objective of QPE is to estimate the eigenphase of a unitary operator. Given a unitary operator $U$ acting on an eigenstate $|\psi\rangle$ satisfying
\[
U|\psi\rangle = e^{2\pi i\phi}|\psi\rangle,
\]
where $\phi \in [0,1)$, the goal of the QPE algorithm is to determine the phase $\phi$ with a desired precision. The ability to efficiently estimate eigenphases makes QPE a central primitive for extracting spectral information from quantum systems and underpins many quantum algorithms of practical interest.

The standard QPE algorithm employs a control register prepared in a uniform superposition, followed by a sequence of controlled applications of increasing powers of the target unitary and an inverse quantum Fourier transform (IQFT) that converts the accumulated phase information into its binary representation \cite{coppersmith1994approximate,Hales}. In the ideal circuit model, this procedure estimates the eigenphase with increasing precision as the number of control qubits increases. Nevertheless, its practical implementation remains challenging because it requires multiple controlled-unitary operations together with a multi-qubit IQFT, both of which contribute significantly to the overall circuit depth and gate complexity. For an $n$-qubit control register, the circuit complexity of the standard implementation scales quadratically with the number of qubits, making experimental realisations increasingly demanding as the system size grows. Recent work has explored low-depth algorithms and improved circuit constructions for phase estimation to address these challenges \cite{Ni2023,Lin2022,Ding2023,Rall2021,mande2023tight}.

Several variants of the QPE algorithm have been proposed to alleviate these experimental challenges. Kitaev's iterative phase estimation algorithm estimates the phase sequentially using a single ancilla qubit, thereby substantially reducing ancillary resources while preserving the fundamental query complexity \cite{Kitaev1995,Cleve1998,Dobsicek2007}. Semiclassical implementations further simplify the inverse quantum Fourier transform by replacing coherent multi-qubit operations with adaptive measurements and classical feedforward \cite{Griffiths1996}. More recently, Bayesian inference techniques, robust phase estimation protocols, and adaptive estimation strategies have been developed to improve noise resilience and reduce the experimental resources required for phase estimation \cite{Wiebe2015,Kimmel2015,Yamamoto2024,Shukla2026,gebhart2021bayesian,svore2013faster}. Randomised and statistical approaches to phase estimation have also been proposed to reduce circuit depth and improve performance in noisy settings \cite{Wan2022,o2019quantum,somma2019quantum}. Additionally, Fourier-transform-based methods and special function techniques have been explored for quantum phase estimation \cite{ChapeauBlondeau2020,hayashi2023special}. Although these approaches considerably simplify particular aspects of the algorithm, they do not fundamentally eliminate the experimental overhead associated with implementing controlled operations, which remains the primary bottleneck for realising scalable QPE on existing quantum hardware.

This challenge is particularly pronounced in photonic quantum computing. Most photonic implementations rely on linear-optical quantum computing architectures, where two-qubit entangling gates are realised probabilistically using the Knill--Laflamme--Milburn (KLM) protocol or its variants. Consequently, controlled-unitary operations are themselves probabilistic, causing the overall success probability of the phase estimation circuit to decrease rapidly as the circuit size increases. Previous photonic demonstrations of QPE have therefore relied on post-selection, probabilistic controlled gates, or adaptive measurement schemes, limiting their scalability despite successful proof-of-principle demonstrations. Developing deterministic photonic realisations of QPE with reduced circuit complexity therefore remains an important open challenge for scalable optical quantum information processing. Hybrid photonic architectures utilizing both polarization and path degrees of freedom have shown promise for implementing multi-qubit protocols with deterministic operations \cite{Wang2018,Souza2022,Sengupta2025,Kolangatt2024}. The photonic quantum walk platform has demonstrated high-fidelity universal quantum gates and scalable entangled-state generation, making it well suited for implementing QPE protocols \cite{Shivani2021,Prateek2023, Sengupta2025, Kolangatt2024}. Beam splitter-based operations for path-encoded qubits have been extensively characterized \cite{Campos1989,Makarov2022}, and Sagnac interferometer configurations provide the phase stability required for deterministic controlled operations \cite{Sagnac1913}.

Fortunately, many practically relevant unitary operators possess additional mathematical structure that need not be treated by the fully general QPE algorithm. In particular, a broad class of diagonal unitary operators composed of tensor products of phase gates with dyadically scaled phases naturally arises in quantum Fourier transform (QFT)-based circuits, cyclic group representations, and periodically evolving quantum systems. The hierarchical phase structure of these operators suggests that the standard QPE circuit may contain substantial redundancy when applied to this important family of unitaries. This naturally raises the following question: can the intrinsic structure of these unitary operators be exploited to simplify quantum phase estimation while simultaneously enabling deterministic photonic implementations? Recent theoretical advances in quantum singular value transformation and related frameworks have provided new perspectives on phase estimation \cite{gilyen2019quantum,martyn2021grand,greenaway2024case}, while unbiased and average-case approaches have extended the applicability of QPE \cite{lu2023unbiased,linden2022average}.

In this work, we answer this question in the affirmative. We introduce a quantum phase estimation protocol specifically tailored to this structured family of diagonal unitary operators, exploiting their hierarchical phase structure to reduce the circuit complexity from $\mathcal{O}(n^2)$ to $\mathcal{O}(n)$. The optimality of quantum clocks and phase estimation has been studied in the context of quantum metrology \cite{buvzek1999optimal,luis1996optimum,rzkadkowski2017discrete,berry2000optimal}, and our approach builds on these fundamental insights. We experimentally demonstrate the protocol on a four-qubit photonic quantum processor. Here, a pair of photons encodes both the two target qubits and the two control qubits by combining their polarization and path degrees of freedom. Unlike previous photonic implementations based on dual-rail encoding and KLM-type architectures, our approach realises the required controlled operations deterministically, thereby eliminating the probabilistic gate overhead that limits existing photonic implementations of QPE. Furthermore, although the protocol is developed for a particular class of structured unitary operators, this class encompasses several practically relevant quantum circuits, making the proposed approach both experimentally attractive and potentially scalable to higher-dimensional systems whenever the underlying unitary structure is preserved. The connection between phase estimation and quantum simulation has been explored extensively \cite{temme2011quantum,somma2019quantum}, and our work contributes to this broader landscape of quantum algorithms for spectral analysis.

The remainder of this paper is organized as follows. In Sec.~II, we introduce the structured family of unitary operators considered in this work and derive the proposed linear-complexity quantum phase estimation protocol. Section~III describes its deterministic photonic implementation using polarization and path encoding. In Sec.~IV, we present the experimental results and compare the performance of the proposed protocol with the conventional implementation. Finally, Sec.~V summarizes our conclusions and discusses future directions.

\section{Linear-Complexity QPE for Structured Unitary Operators}

Quantum phase estimation (QPE) is designed to estimate the eigenphase of a unitary operator. Consider a unitary operator $U$ and one of its eigenvectors $|\psi\rangle$ satisfying
\begin{equation}
U|\psi\rangle = e^{2\pi i\phi}|\psi\rangle,
\end{equation}
where $\phi\in[0,1)$ denotes the eigenphase. The objective of the QPE algorithm is to estimate $\phi$ with a desired precision by encoding the phase information into a control quantum register. The standard QPE algorithm employs two quantum registers: a control (or clock) register and a target register. The target register stores the eigenstate $|\psi\rangle$, while the control register is used to encode the binary representation of the eigenphase. The number of qubits in the control register determines both the achievable precision and the success probability of the phase estimation protocol. The conventional QPE protocol consists of three stages. Consider a control register consisting of $m$ qubits and a target register consisting of $n$ qubits. In the first stage, Hadamard gates are applied to each qubit of the control register, creating an equal superposition of all computational basis states. This is followed by a sequence of controlled-unitary operations, where the $i^{\mathrm{th}}$ control qubit applies the operator $U^{2^{i}}$ on the target register, with $i=0,1,\ldots,m-1$. These controlled operations exploit the phase-kickback mechanism, transferring the eigenphase of the target state onto the control register while leaving the target register unchanged. Consequently, the control register evolves to

\begin{equation}
\label{eq:qpe_binary_rep}
\begin{aligned}
\frac{1}{2^{m/2}}
&
\left(|0\rangle+e^{2\pi i2^{m-1}\phi}|1\rangle\right)
\cdots
\left(|0\rangle+e^{2\pi i2^{0}\phi}|1\rangle\right)
\\
&=
\frac{1}{2^{m/2}}
\left(|0\rangle+e^{2\pi i0.\phi_m}|1\rangle\right)
\left(|0\rangle+e^{2\pi i0.\phi_{m-1}\phi_m}|1\rangle\right)
\\
&
\qquad\cdots
\left(|0\rangle+e^{2\pi i0.\phi_1\phi_2\cdots\phi_m}|1\rangle\right)
\\
&=
\frac{1}{2^{m/2}}
\sum_{x=0}^{2^m-1}
e^{2\pi i\phi_x}|x\rangle.
\end{aligned}
\end{equation}

where the eigenphase admits the $m$-bit binary representation
\[
\phi = 0.\phi_1\phi_2\cdots\phi_m.
\]

Equation~(\ref{eq:qpe_binary_rep}) clearly illustrates that the binary digits of the eigenphase are encoded coherently across the control register through phase kickback, while the target register remains in the eigenstate $|\psi\rangle$ throughout the evolution.

The second stage of the protocol applies the inverse Quantum Fourier Transform (IQFT) to the control register. The role of the IQFT is to convert the accumulated phase information into the computational basis, thereby recovering the binary representation of the eigenphase. The resulting transformation is

\begin{equation}
\label{eq:2}
\begin{aligned}
&
\frac{1}{2^{m/2}}
\left(|0\rangle_m+e^{2\pi i(0.\phi_m)}|1\rangle_m\right)
\otimes\cdots
\\
\otimes&
\left(|0\rangle_1+e^{2\pi i(0.\phi_1\phi_2\cdots\phi_m)}|1\rangle_1\right)
\xrightarrow{\mathrm{IQFT}}
|\phi_1\phi_2\cdots\phi_m\rangle.
\end{aligned}
\end{equation}

Finally, the control register is measured in the computational basis to obtain an estimate of the eigenphase. When $\phi$ is exactly representable as an integer multiple of $1/2^m$, the measurement yields the exact binary representation of the phase. Otherwise, the measurement outcomes follow a probability distribution that is sharply peaked around the true eigenphase, allowing an accurate estimate to be obtained with high probability.

Although the standard QPE algorithm is theoretically efficient, its practical implementation becomes increasingly demanding as the number of qubits increases. The principal source of this complexity is the inverse Quantum Fourier Transform together with the sequence of controlled-unitary operations, resulting in an overall circuit complexity that scales as $\mathcal{O}(n^2)$. This overhead becomes particularly significant in experimental platforms, such as linear-optical quantum computing, where controlled operations constitute the dominant experimental resource.

In this work, we show that this complexity can be substantially reduced for an important class of structured diagonal unitary operators of the form
\begin{equation}
\label{eq:4}
U = P(\pi)\otimes P\!\left(\frac{\pi}{2}\right)\otimes \cdots \otimes P\!\left(\frac{\pi}{2^{n-2}}\right)\otimes P\!\left(\frac{\pi}{2^{n-1}}\right)
\end{equation}

where $P(\theta)$ denotes a single-qubit phase gate. Such unitary operators naturally arise in quantum Fourier transform-based circuits and related quantum information processing tasks. Their hierarchical phase structure enables a significant simplification of the phase estimation procedure. As we show in the following sections, this structure allows the IQFT stage to be eliminated entirely, reducing the overall circuit complexity from $\mathcal{O}(n^2)$ to $\mathcal{O}(n)$ while preserving the functionality of the standard QPE algorithm. This reduction forms the basis of the deterministic photonic implementation presented in this work.

\subsection{Standard QPE Algorithm}

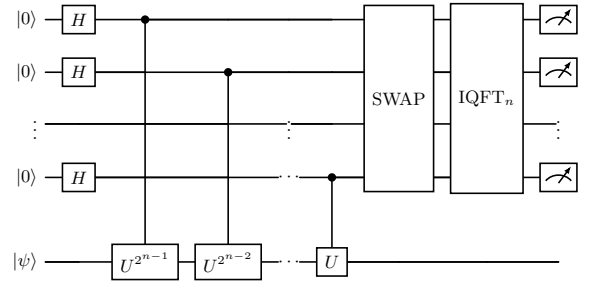
\begin{figure}[h!]
\centering
\scalebox{0.75}{%
\begin{quantikz}[row sep=0.4cm, column sep=0.3cm]
\lstick{$|0\rangle$} & \gate{H} & \ctrl{4} & \qw & \qw & \qw
    & \gate[4,nwires=3]{\text{SWAP}} & \gate[4,nwires=3]{\text{IQFT$_n$}}
    & \meter{} \\
\lstick{$|0\rangle$} & \gate{H} & \qw & \ctrl{3} & \qw & \qw
    & \qw & \qw & \meter{} \\
\lstick{$\vdots$} & \qw & \qw & \qw & \vdots & \qw
    & & & \vdots \\
\lstick{$|0\rangle$} & \gate{H} & \qw & \qw & \cdots & \ctrl{1}
    & \qw & \qw & \meter{} \\[0.5cm]
\lstick{$|\psi\rangle$} & \qw
    & \gate{U^{2^{n-1}}} & \gate{U^{2^{n-2}}} & \cdots & \gate{U}
    & \qw & \qw & \qw \\
\end{quantikz}
}%
\caption{Computational circuit for the standard Quantum Phase Estimation algorithm. The controlled-$U^{2^k}$ operations are applied sequentially, followed by a swap network (reversed order) and the inverse quantum Fourier transform.}
\label{fig:qpe}
\end{figure}

Here, we present the standard approach for the QPE algorithm for the class of structured unitary operators in tensor product form, as given in Eq.~\eqref{eq:4}. Each $P(\theta)$ represents the phase gate acting on a target qubit. The computational circuit implementing the standard QPE with this unitary is shown in Fig.~\ref{fig:qpe}. The circuit consists of three main stages: (i) creation of a uniform superposition over the control register, (ii) application of controlled-$U$ operations, and (iii) an IQFT followed by measurement.

The standard QPE procedure for our specific unitary can be described step by step as follows. Let the target register be in the computational basis state $|c\rangle = |c_1 c_2 \cdots c_n\rangle$, with $c_i \in \{0,1\}$. The algorithm proceeds as shown below.\\
\\

\noindent\textbf{Algorithm 1.} Standard Quantum Phase Estimation for
$U=P(\pi)\otimes P(\pi/2)\otimes\cdots\otimes P(\pi/2^{n-1})$

\vspace{0.5em}

\noindent\textbf{Input:} Control register
$|0\rangle^{\otimes n}$, Target register
$|c_1c_2\cdots c_n\rangle$.

\noindent\textbf{Output:} Recovered state
$|c_1c_2\cdots c_n\rangle\otimes|c_1c_2\cdots c_n\rangle$.

\vspace{0.8em}

\noindent\textbf{Step 1:} \textit{Initial State}

\[
|0\rangle^{\otimes n}|c_1c_2\cdots c_n\rangle
\]

\noindent\textbf{Step 2:} \textit{Apply Hadamard Gates}

\begin{gather*}
\rightarrow
\frac{1}{\sqrt{2^n}}
\sum_{x=0}^{2^n-1}|x\rangle|c\rangle,\\
x=x_1\otimes x_2\otimes\cdots\otimes x_n,\qquad
c=c_1\otimes c_2\otimes\cdots\otimes c_n.
\end{gather*}

\noindent\textbf{Step 3:} \textit{Apply Controlled Unitaries}

\[
\rightarrow
\frac{1}{\sqrt{2^n}}
\sum_{x=0}^{2^n-1}
e^{i\Phi(x)}
|x\rangle|c\rangle,
\]

\noindent where the accumulated phase is

\begin{multline*}
\Phi(x)=
x_1\!\left(\pi c_n+\frac{\pi}{2}c_{n-1}
+\cdots+\frac{\pi}{2^{n-1}}c_1\right)\\
+x_2\!\left(\pi c_n+\frac{\pi}{2}c_{n-1}
+\cdots+\frac{\pi}{2^{n-2}}c_2\right)\\
+x_3\!\left(\pi c_n+\frac{\pi}{2}c_{n-1}
+\cdots+\frac{\pi}{2^{n-3}}c_3\right)\\
+\cdots+x_n(\pi c_n).
\end{multline*}

\noindent\textbf{Step 4:} \textit{Factorize the Phase Term}

\[
\begin{aligned}
\rightarrow\;&
\frac{1}{\sqrt{2^n}}
\Bigl(|0\rangle+e^{2\pi i(0.c_n)}|1\rangle\Bigr)\otimes\\[1mm]
&
\Bigl(|0\rangle+e^{2\pi i(0.c_{n-1}c_n)}|1\rangle\Bigr)\otimes
\cdots\otimes\\[1mm]
&
\Bigl(|0\rangle+e^{2\pi i(0.c_1c_2\cdots c_n)}|1\rangle\Bigr)
|c\rangle.
\end{aligned}
\]

\noindent\textbf{Step 5:} \textit{Apply SWAP Gates (Reverse Order)}

\[
\begin{aligned}
\rightarrow\;&
\frac{1}{\sqrt{2^n}}
\Bigl(|0\rangle+e^{2\pi i(0.c_1c_2\cdots c_n)}|1\rangle\Bigr)\otimes\\[1mm]
&
\Bigl(|0\rangle+e^{2\pi i(0.c_2c_3\cdots c_n)}|1\rangle\Bigr)\otimes
\cdots\otimes\\[1mm]
&
\Bigl(|0\rangle+e^{2\pi i(0.c_n)}|1\rangle\Bigr)
|c\rangle.
\end{aligned}
\]

\noindent\textbf{Step 6:} \textit{Recognize the QFT State}

The state obtained in Step~5 is exactly the quantum Fourier transform of the computational basis state
$|c_1c_2\cdots c_n\rangle$:

\[
\rightarrow
\mathrm{QFT}\,
|c_1c_2\cdots c_n\rangle
|c\rangle.
\]

\noindent\textbf{Step 7:} Apply the Inverse QFT

\[
\rightarrow
|c_1c_2\cdots c_n\rangle\otimes
|c_1c_2\cdots c_n\rangle.
\]

\noindent\textbf{Final Output}

\[
|c_1c_2\cdots c_n\rangle\otimes
|c_1c_2\cdots c_n\rangle.
\]

This derivation shows that the standard QPE circuit, when applied to our structured unitary, successfully copies the target state onto the control register. The crucial observation is that the phase accumulated in Step 3 factors exactly into the binary fractions that appear in the QFT basis states, making the subsequent IQFT step necessary to recover the original computational basis labels.

\subsection{Optimised Computational Scheme}

A careful inspection of the standard QPE circuit reveals an opportunity for significant simplification. The structure of the unitary $U$ - specifically, the fact that its eigenvalues are already encoded as phase factors of the form $e^{2\pi i (0.c_k \cdots c_n)}$ - allows us to bypass the IQFT stage entirely. Instead of performing the full QPE procedure with controlled-$U^{2^k}$ gates followed by an IQFT, we can implement the same state transformation using only Hadamard gates and controlled-$Z$ operations.

\begin{figure}[h]
\centering
\scalebox{0.7}{
\begin{quantikz}
\lstick{$|0\rangle$} & \gate{H} & \ctrl{4} & \qw      & \qw      & \qw      & \gate{H} & \meter{} \\
\lstick{$|0\rangle$} & \gate{H} & \qw      & \ctrl{4} & \qw      & \qw      & \gate{H} & \meter{} \\
\vdots               &          &          &          &          &          & \vdots   &          \\
\lstick{$|0\rangle$} & \gate{H} & \qw      & \qw      & \qw \dots\dots      & \ctrl{4} & \gate{H} & \meter{} \\[0.5cm]
\lstick{$|\psi\rangle$} & \qw   & \gate{Z} & \qw      & \qw      & \qw      & \qw      & \qw      \\
\lstick{$|\psi\rangle$} & \qw   & \qw      & \gate{Z} & \qw      & \qw      & \qw      & \qw      \\
\vdots               &          &          &          &          &          &          &          \\
\lstick{$|\psi\rangle$} & \qw   & \qw      & \qw      & \qw   \dots \dots   & \gate{Z} & \qw      & \qw      \\
\end{quantikz}
}
\caption{Optimised Quantum Phase Estimation Circuit for $n$ control qubits.}
\label{fig:optimised}
\end{figure}
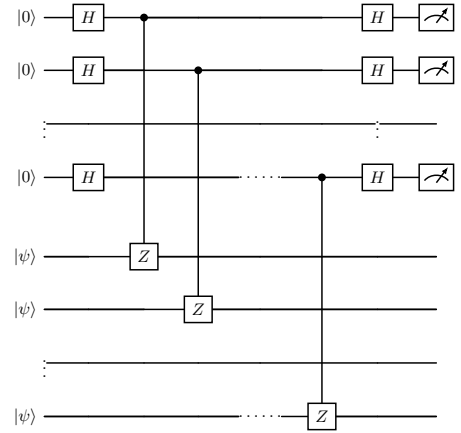

The optimised circuit is shown in Fig.~\ref{fig:optimised}. It replaces the controlled-$U^{2^k}$ gates with controlled-$Z$ gates acting on each control qubit, and eliminates the SWAP network and the IQFT entirely. This reduces the circuit complexity from $\mathcal{O}(n^2)$ to $\mathcal{O}(n)$ gates, while preserving the functionality of the original implementation. Moreover, the scheme is deterministic, achieving the desired output state without any post-selection or measurement feedback.

The state evolution of the optimised circuit is as follows. Starting from the initial state $|0\rangle^{\otimes n} \otimes |c\rangle$, we apply Hadamard gates to the control register, then a sequence of controlled-$Z$ gates, and finally another round of Hadamard gates.The procedure is outlined below.\\

\noindent\textbf{Algorithm 2.} Optimised Computational Scheme Using Hadamard and Controlled-Z Gates

\vspace{0.5em}

\noindent\textbf{Input:} Control register
$|0\rangle^{\otimes n}$, Target register
$|c_1c_2\cdots c_n\rangle$.

\noindent\textbf{Output:}
$|c_1c_2\cdots c_n\rangle\otimes|c_1c_2\cdots c_n\rangle$.

\vspace{0.8em}

\noindent\textbf{Step 1:} Initial State
\[
|0\rangle^{\otimes n}\otimes|c_1c_2\cdots c_n\rangle
\]

\noindent\textbf{Step 2:} Apply Hadamard Gates

\begin{gather*}
\rightarrow
\frac{1}{\sqrt{2^n}}
\sum_{x=0}^{2^n-1}|x\rangle|c\rangle,\\
x=x_1\otimes x_2\otimes\cdots\otimes x_n,\qquad
c=c_1\otimes c_2\otimes\cdots\otimes c_n.
\end{gather*}

\noindent\textbf{Step 3:} Apply Controlled-$Z$ Gates
\[
\rightarrow
\frac{1}{\sqrt{2^n}}
\sum_{x=0}^{2^n-1}
(-1)^{x\cdot c}
|x\rangle\otimes|c\rangle,
\]
\noindent where $x\cdot c = x_1c_1 + x_2c_2 + \cdots + x_nc_n$.

\vspace{6pt}
\noindent\textbf{Step 4:} Apply Hadamard Gates Again

\[
\begin{aligned}
\rightarrow\;&
\frac{1}{2^n}
\sum_{x=0}^{2^n-1}
\sum_{y=0}^{2^n-1}
(-1)^{x\cdot c}
(-1)^{x\cdot y}
|y\rangle\otimes|c\rangle\\[2mm]
=\;&
\frac{1}{2^n}
\sum_{\substack{x_i=0\\y_i=0}}^{1}
(-1)^{\sum_{i=1}^{n}x_i(c_i+y_i)}
|y_1\cdots y_n\rangle
\otimes
|c_1\cdots c_n\rangle.
\end{aligned}
\]

Using the identity
\[
\sum_{x_i=0}^{1}
(-1)^{x_i(c_i+y_i)}
=
2\delta_{c_i,y_i},
\]

the sum over $x$ collapses, yielding
\[
\rightarrow
|c_1c_2\cdots c_n\rangle
\otimes
|c_1c_2\cdots c_n\rangle.
\]

\noindent\textbf{Final Output}

\[
|c_1c_2\cdots c_n\rangle
\otimes
|c_1c_2\cdots c_n\rangle.
\]

The equivalence of the standard and optimised schemes is illustrated in Fig.~\ref{fig:comb} for the case $n=2$. The optimised circuit produces exactly the same output state as the standard QPE, confirming that the simplification preserves the desired functionality while reducing the overall gate count and removing the need for the IQFT.
The optimised computational scheme thus provides a more efficient, scalable, and deterministic implementation of quantum state copying for the considered class of unitaries. By exploiting the special structure of the phase gates, we have effectively replaced the full QPE protocol with a linear-depth circuit that achieves the same objective, paving the way for practical applications in quantum information processing.

\vspace{6pt}
\section{Experimental Implementation using Photonic Qubits}

\subsection{Experimental Implementation}

The proposed protocol is experimentally implemented on a photonic quantum processor based on the quantum walk architecture for universal quantum computation \cite{Sengupta2025,Kolangatt2024}. This platform is particularly well-suited for our scheme, as it naturally supports the structured unitary operations central to our QPE protocol. The experimental setup utilises the hybrid encoding capabilities of the photonic quantum walk architecture. The polarization states, $|H\rangle$ and $|V\rangle$, represent one qubit, while additional qubits are encoded in the spatial path degrees of freedom generated by a network of beam splitters \cite{Sengupta2025,Kolangatt2024}. This hybrid encoding enables multiple qubits to be manipulated using a single photon, significantly reducing the need for multi-photon interference and thereby substantially lowering the physical photon resources required for multi-qubit operations. Moreover, the deterministic nature of the gate operations in this architecture eliminates the post-selection overhead typically associated with linear optical quantum computing. Universal single-qubit operations are realised using combinations of quarter-wave plates (QWPs) and half-wave plates (HWPs), providing complete control over the polarization qubit. Deterministic controlled operations between polarization and path qubits are implemented using the photonic quantum walk architecture \cite{Sengupta2025,Kolangatt2024}, which allows for the realisation of entangling gates with high fidelity. The beamsplitter network generates the spatial modes required for encoding multiple path qubits, while the quantum walk dynamics provide the necessary interactions for implementing controlled unitary operations. Multiple-qubit quantum state preparation and computation based on these internal degrees of freedom of photons have already been demonstrated \cite{Wang2018,Souza2022,Sengupta2025,Kolangatt2024}. In particular, the platform has previously demonstrated high-fidelity realisation of universal quantum gates, four- and six-qubit entangled-state generation, making it well 
\newpage
\onecolumngrid

\begin{figure*}[t]
\centering
\includegraphics[width=\textwidth]{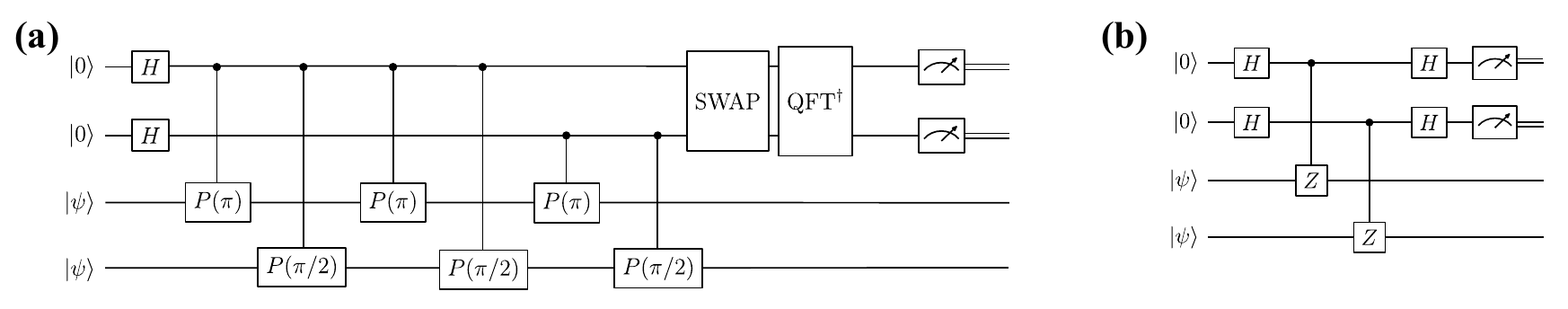}
\caption{(a) Standard circuit for the \(n=2\) quantum phase estimation algorithm with the unitary \(U=P(\pi)\otimes P(\pi/2)\). (b) The optimised circuit corresponding to (a).}
\label{fig:comb}
\end{figure*}

\twocolumngrid
suited for implementing the multi-qubit quantum information processing protocols described in this work. The combination of high-fidelity single-qubit operations and deterministic controlled gates ensures that our QPE implementation benefits from both the reduced resource
requirements and the operational reliability of the photonic quantum walk platform. The experimental feasibility of our optimised QPE scheme is thus supported by the demonstrated capabilities of the photonic quantum walk architecture. The ability to encode multiple qubits in a single photon, combined with the deterministic implementation of controlled operations, directly addresses the scalability challenges faced by conventional photonic quantum computing approaches. This makes the photonic quantum walk platform an ideal testbed for verifying the theoretical predictions of our optimised QPE protocol and for exploring its potential applications in practical quantum information processing tasks.

\section{Source, Circuit Implementation, and Experimental Results} 

 \subsection{Experimental Implementation}

\begin{figure*}[t]
\centering
\includegraphics[width=\textwidth]{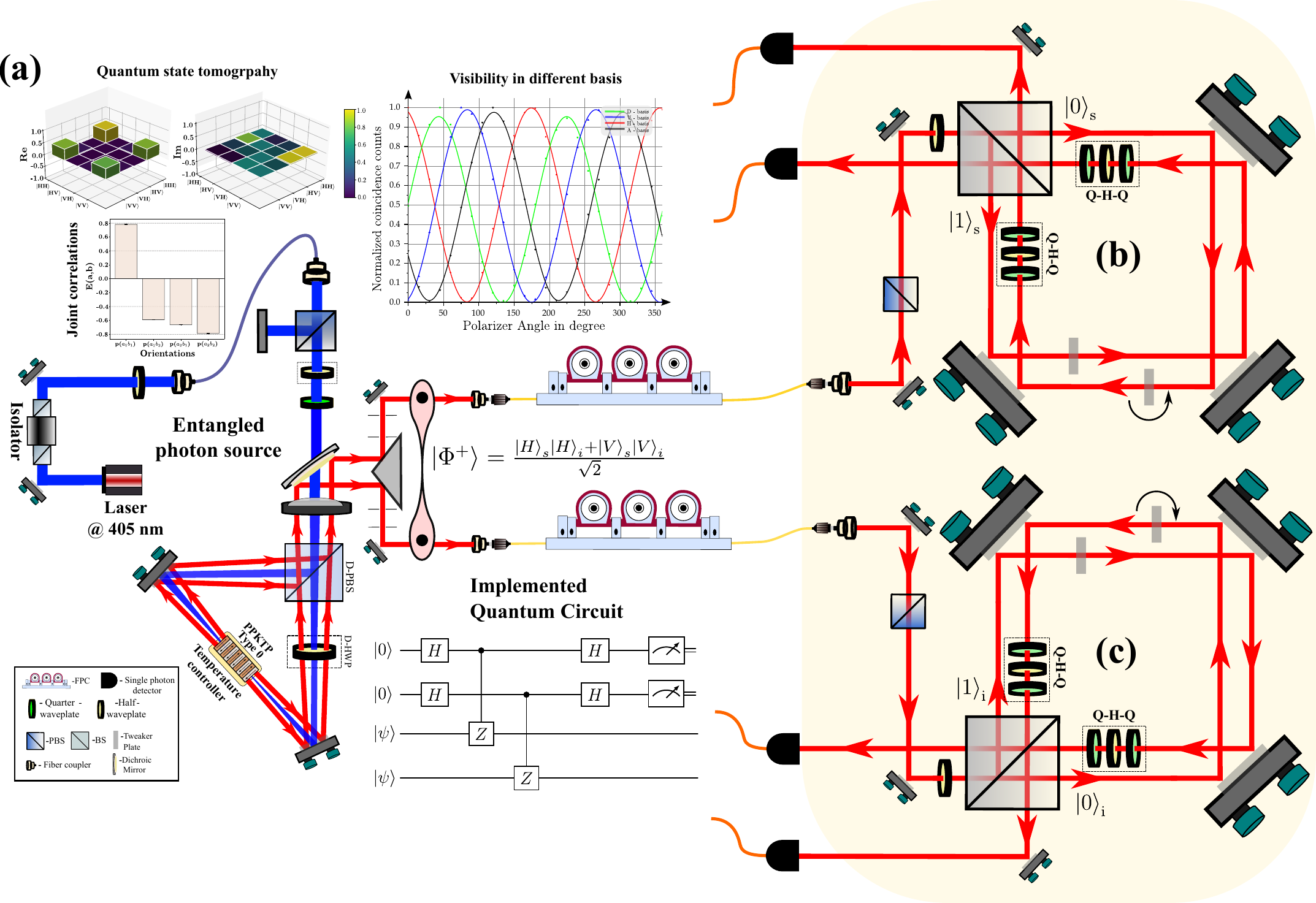}
\caption{Experimental setup for the implementation of the two-control, two-target quantum phase estimation algorithm (\(n=2\)) with the unitary \(U=P(\pi)\otimes P(\pi/2)\) using path- and polarization-encoded photonic qubits. (a) Type-0 polarization-entangled photon source and its characterisation. The source exhibits visibilities of 98.9\% and 98.1\% in the H/V and A/D bases, respectively, violates the CHSH inequality with \(S=2.72\pm0.03\), and quantum state tomography confirms the generation of a Bell state. (b), (c) Displaced Sagnac interferometers encoding the first and third qubits, and the second and fourth qubits, respectively. The average single-photon interference visibility measured in the two interferometers was approximately 93\%.}
\label{fig:Expt_schematic}
\end{figure*}

We experimentally realised a two-qubit QPE circuit comprising four qubits, using photons entangled in polarization and path degrees of freedom. The chosen unitary operator is given by $U = P(\pi) \otimes P(\pi/2)$, where $P(\theta)$ is the phase gate. In matrix form, this unitary is expressed as
\[
U=P(\pi)\otimes P\!\left(\frac{\pi}{2}\right)
=
\begin{pmatrix}
1 & 0 & 0 & 0\\
0 & i & 0 & 0\\
0 & 0 & -1 & 0\\
0 & 0 & 0 & -i
\end{pmatrix}.
\]
The standard two-qubit QPE circuit corresponding to the unitary operator \(U\) is shown in Fig.~\ref{fig:comb}(a), while the equivalent optimised circuit is presented in Fig.~\ref{fig:comb}(b). The experimental realisation of the optimised scheme using photonic qubits is illustrated in Fig.~\ref{fig:Expt_schematic}.

To generate single photons, we employed a Type-0 spontaneous parametric down-conversion (SPDC) source based on a Sagnac-type interferometer. The nonlinear crystal used is a periodically poled potassium titanyl phosphate (PPKTP) crystal with a poling period of $3.425\,\mu\text{m}$. The source generates photon pairs entangled in polarization, with the two-photon state given by
\[
\frac{1}{\sqrt{2}}\bigl(|H\rangle_s |H\rangle_i + |V\rangle_s |V\rangle_i\bigr),
\]
where $|H\rangle$ and $|V\rangle$ denote horizontal and vertical polarization, respectively. The quality of the entangled source was characterised through multiple measurements. The visibility in the H/V and A/D bases was measured to be $98.9\%$ and $98.1\%$, respectively. Furthermore, the CHSH inequality violation was calculated to be $S = 2.72 \pm 0.03$, significantly exceeding the classical bound of 2. These characterisation results confirm the high-quality polarization entanglement of the source, making it well-suited for implementing multi-qubit quantum information protocols.

The control qubits (qubits 1 and 2 in Fig.~\ref{fig:comb}(b)) are encoded in the path degree of freedom of the photons, while the target qubits (qubits 3 and 4) are encoded in the polarization degree of freedom. The eigenvectors corresponding to the eigenvalues $1$, $i$, $-1$, and $-i$ of the unitary $U$ are $|HH\rangle$, $|HV\rangle$, $|VH\rangle$, and $|VV\rangle$, respectively, where $|H\rangle$ is encoded as $|0\rangle$ and $|V\rangle$ is encoded as $|1\rangle$. A Hadamard operation on the path qubit is realised using a 50:50 beam splitter \cite{Campos1989,Makarov2022}, which creates a superposition of the two paths. The first and last Hadamard operations on the path qubit, implemented using beam splitters, form an interferometric setup. The controlled-$Z$ gate between the path and polarization qubits is implemented by placing a quarter-wave plate (QWP), half-wave plate (HWP), and quarter-wave plate (QWP) combination in the reflected arm of the interferometer \cite{Sengupta2025}. Thus, each control–target qubit pair forms an independent interferometer.

The experimental implementation of the optimised computational scheme is shown in Fig.~\ref{fig:Expt_schematic}. The displaced Sagnac interferometer shown in Fig.~\ref{fig:Expt_schematic}(b) realises the the first control-target qubit pair (qubits 1 and 3), while the interferometer shown in Fig.~\ref{fig:Expt_schematic}(c) realises the second control-target qubit pair (qubits 2 and 4). The average single-photon interference visibility measured in 
\newpage
\onecolumngrid

\twocolumngrid

the two interferometers was approximately \(93\%\). The experimental data are obtained from coincidence measurements between the photons emerging from the two interferometers, enabling the reconstruction of the output state and the verification of the quantum phase estimation protocol. Each photon simultaneously encodes one control qubit in its path degree of freedom and one target qubit in its polarization degree of freedom, and the first control qubit acts exclusively on the first target qubit, the second control qubit acts exclusively on the second target qubit, and so on, establishing a one-to-one correspondence between the control and target registers.
Consequently, each control-target qubit pair is realised as an independent displaced Sagnac interferometer. This architecture naturally extends to an \(n\)-control, \(n\)-target implementation using \(n\) photons and \(n\) independent interferometers. As no interactions are required between different control-target qubit pairs, the implementation remains deterministic while preserving the linear scaling of the optimised computational architecture.

The experimental platform thus successfully implements the optimised two-qubit QPE circuit using hybrid path-polarization encoding of photonic qubits. The high-quality entangled photon source, combined with the robust displaced-Sagnac interferometers, provides a practical and scalable approach to realising the deterministic controlled operations central to our protocol. The experimental results validate the theoretical predictions of the optimised QPE scheme, demonstrating that the simplified circuit preserves the functionality of the standard implementation while significantly reducing the resource requirements and circuit complexity.

\subsection{Experimental Results}

\begin{figure*}[t]
\centering
\includegraphics[width=\textwidth]{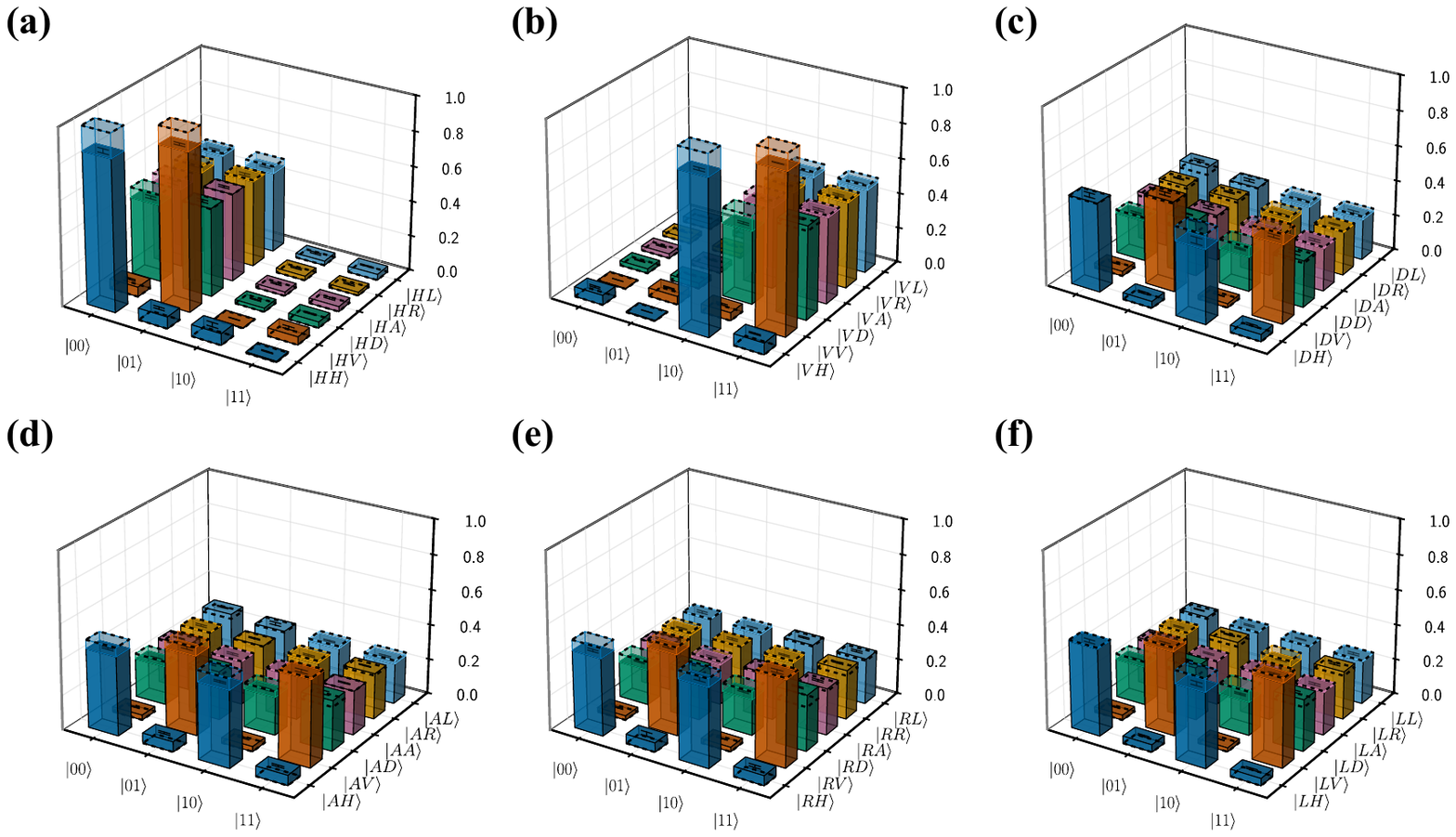}
\caption{Comparison of the experimentally measured and theoretically expected normalised coincidence counts for the complete set of 36 polarization projection states. The solid bars represent the experimentally measured coincidence counts, while the overlaid dotted bars denote the corresponding theoretical predictions. Panels (a) to (f) correspond to measurements with the first photon's polarization projected onto the states \(\ket{H}\), \(\ket{V}\), \(\ket{D}\), \(\ket{A}\), \(\ket{R}\), and \(\ket{L}\), respectively, while the second photon's polarization is projected onto the complete basis \(\{\ket{H},\ket{V},\ket{D},\ket{A},\ket{R},\ket{L}\}\). The close overlap between the experimental and theoretical coincidence counts demonstrates good agreement with the expected outcomes.}
\label{fig:final}
\end{figure*}

The experimental results are presented in Fig.~\ref{fig:final}. To verify the implementation of the unitary operator 
\newpage
\onecolumngrid

\twocolumngrid
$U = P(\pi)\otimes P(\pi/2)$, normalised coincidence counts were measured for all two-qubit combinations of the polarization states $|H\rangle$, $|V\rangle$, $|D\rangle$, $|A\rangle$, $|R\rangle$, and $|L\rangle$. These states constitute the complete eigenbases of the Pauli operators $\sigma_z$, $\sigma_x$, and $\sigma_y$, satisfying the completeness relation
\[
I=\sum_i |\psi_i\rangle\langle\psi_i|.
\]
The input states were prepared using a polarizing beam splitter (PBS), together with a half-wave plate (HWP) or quarter-wave plate (QWP), at the input of each interferometer, as shown in Fig.~\ref{fig:Expt_schematic}. For each prepared state, the coincidence counts at the four output ports were recorded and normalised.

Figure~\ref{fig:final} shows the measured coincidence distributions for the six polarization bases. The bars correspond to the experimentally measured normalised coincidence counts, while the theoretical predictions are overlaid for comparison. The measurements demonstrate that a single dominant coincidence peak is obtained only when the prepared input state is an eigenvector of the implemented unitary operator $U = P(\pi)\otimes P(\pi/2)$. In contrast, when the input state is not an eigenvector, the coincidence counts are distributed among multiple output ports according to the projection of the input state onto the eigenbasis of the unitary.

The computational basis states \(\ket{00}\), \(\ket{01}\), \(\ket{10}\), and \(\ket{11}\), corresponding to the eigenvectors of the unitary operator \(U=P(\pi)\otimes P(\pi/2)\), are encoded as the polarization states \(\ket{HH}\), \(\ket{HV}\), \(\ket{VH}\), and \(\ket{VV}\), respectively. Consequently, the dominant coincidence peaks are observed for the projections onto \(\ket{HH}\), \(\ket{HV}\), \(\ket{VH}\), and \(\ket{VV}\), corresponding to the computational basis states \(\ket{00}\), \(\ket{01}\), \(\ket{10}\), and \(\ket{11}\), respectively. In contrast, measurements performed in the diagonal/antidiagonal and right/left circular polarization bases yield the expected superposition of coincidence counts, as these states are not eigenvectors of the implemented unitary. The close agreement between the experimentally measured and theoretically expected coincidence distributions across all 36 projection measurements confirms the correct operation of the implemented quantum phase estimation algorithm.

The experimental results demonstrate the successful estimation of the eigenphases associated with the eigenvectors of the unitary operator \(U=P(\pi)\otimes P(\pi/2)\). For each input eigenvector, the quantum phase estimation algorithm maps the corresponding eigenvalue to a unique computational basis state of the control register, resulting in a distinct coincidence signature at the output. Specifically, the eigenvectors \(\ket{00}\), \(\ket{01}\), \(\ket{10}\), and \(\ket{11}\), corresponding to the eigenvalues \(1\), \(i\), \(-1\), and \(-i\), respectively, produce the expected output distributions, thereby identifying the associated eigenphases. Evaluated at a 2-bit precision level ($m=2$), these eigenvalues correspond to the exact binary eigenphases $\phi = 0.00, 0.01, 0.10,$ and $0.11$ (corresponding to values of $0$,$1/4$, $1/2$, and $3/4$). Because each phase is an exact integer multiple of $1/2^m$, the system produces completely deterministic output distributions, thereby identifying the associated eigenphases without phase leakage.

The experimental results are consistent with the expected outcomes. The non-zero coincidence counts observed in the undesired output ports are attributable to experimental imperfections, including source non-idealities, waveplate calibration errors, and interferometer instabilities. Despite these imperfections, the clear distinction between the dominant and off-diagonal coincidence peaks provides unambiguous evidence for the successful implementation of the optimised QPE protocol. The agreement between theory and experiment demonstrates the validity of the theoretical framework and the practical feasibility of the optimised computational scheme, paving the way for its application to larger-scale quantum information processing tasks.

\section{Conclusion}

%



In this work, we have proposed an optimised quantum phase estimation (QPE) algorithm tailored for a specific class of structured unitary operators, reducing the gate complexity from $\mathcal{O}(n^2)$ to $\mathcal{O}(n)$. Taking advantage of the inherent structure of the unitary operator, the standard QPE circuit is simplified by eliminating the inverse quantum Fourier transform and replacing controlled operations $U^{2^k}$ with controlled $Z$ gates, significantly reducing experimental resource requirements while preserving the functionality of the original algorithm.

Using a hybrid photonic platform, based on the quantum walk architecture for universal quantum computation, we have experimentally validated the theoretical framework.  Mapping the control register to the path degree of freedom and the target qubits to the polarisation degree of freedom helped decompose the circuit into $n$ independent, modular interferometers.The experimental results reported for $n=2$ show excellent agreement with theoretical results across all 36 projection measurements.

In conclusion, this demonstrates that structural simplifications in quantum algorithms can yield substantial practical benefits by reducing both circuit complexity and the experimental overhead required for implementation. The successful experimental realisation of the optimized QPE protocol using photonic qubits validates the theoretical framework and highlights the potential of hybrid photonic architectures for scalable quantum information processing. We anticipate that the optimization strategy presented here will find applications in a variety of quantum algorithms, contributing to the ongoing efforts to make quantum computing more accessible and practical. Ultimately, this offers a practical path toward scalable photonic quantum information processing.

\begin{acknowledgments}
We thank Ms. Nathar Rafiya Noorie for helping us in setting up a Sagnac interferometer. We acknowledge funding support from the National Quantum Mission, an initiative of the Department of Science and Technology, Govt. of India.
\end{acknowledgments}


\end{document}